\documentstyle[12pt]{article}
\def\ga{\gamma}
\def\de{\delta}

\def\ps{\psi}
\def\om{\omega}

\def\De{\Delta}

\def\cE{{\cal E}}
\def\fr#1#2{{{#1} \over {#2}}}

\def\frac#1#2{{\textstyle{{#1}\over {#2}}}}

\def\lsim{\mathrel{\rlap{\lower4pt\hbox{\hskip1pt$\sim$}}
    \raise1pt\hbox{$<$}}}
\def\gsim{\mathrel{\rlap{\lower4pt\hbox{\hskip1pt$\sim$}}
    \raise1pt\hbox{$>$}}}
\def\sqr#1#2{{\vcenter{\vbox{\hrule height.#2pt
         \hbox{\vrule width.#2pt height#1pt \kern#1pt
         \vrule width.#2pt}
         \hrule height.#2pt}}}}

\newcommand{\beq}{\begin{equation}}
\newcommand{\eeq}{\end{equation}}
\newcommand{\bea}{\begin{eqnarray}}
\newcommand{\eea}{\end{eqnarray}}
\newcommand{\rf}[1]{(\ref{#1})}
 
\renewenvironment{thebibliography}[1]
 { \rm
   \begin{list}{\arabic{enumi}.}
    {\usecounter{enumi} \setlength{\parsep}{0pt}
     \setlength{\itemsep}{3pt} \settowidth{\labelwidth}{#1.}
     \sloppy
    }}{\end{list}}
 
\begin{document}
\titlepage
 
\begin{flushright}
{IUHET 358\\}
{COLBY-97-01\\}
{March 1997\\}
\end{flushright}

\vglue 1cm
	    
\begin{center}
{{\bf TESTING CPT WITH ANOMALOUS MAGNETIC MOMENTS\\}
\vglue 1.0cm
{Robert Bluhm,$^a$ V. Alan Kosteleck\'y,$^b$ and
Neil Russell$^b$\\} 
\bigskip
{\it $^a$Physics Department, Colby College\\}
\medskip
{\it Waterville, ME 04901, U.S.A.\\}
\bigskip
{\it $^b$Physics Department, Indiana University\\}
\medskip
{\it Bloomington, IN 47405, U.S.A.\\}
}
\vglue 0.8cm
 
\end{center}
 
{\rightskip=3pc\leftskip=3pc\noindent
A theoretical framework is introduced 
that describes possible CPT-violating effects in the
context of quantum electrodynamics.
Experiments comparing the anomalous magnetic moments
of the electron and the positron
can place tight limits on CPT violation.
The conventional figure of merit adopted in these experiments,
involving the difference between the corresponding $g$ factors, 
is shown to provide a misleading measure
of the precision of CPT limits.
We introduce an alternative figure of merit,
comparable to one commonly used
in CPT tests with neutral mesons. 
To measure it,
a straightforward extension of current
experimental procedures is proposed.
With current technology,
a CPT bound better than about one part in $10^{20}$ is attainable.

}

\vskip 1 cm


\vskip 1.2cm
\centerline{\it Accepted for publication in Physical Review Letters} 
\medskip
\centerline{\it Scheduled for the issue of August 18 1997}

\newpage
 
\baselineskip=20pt

The CPT theorem 
\cite{pct}
is a powerful result
holding for local relativistic quantum field theories of
point particles in flat spacetime.
It states that such theories
must be invariant under the combined operations
of charge conjugation C,
parity reversal P,
and time reversal T.
Among the implications of the theorem are the equality
of particle and antiparticle masses and lifetimes.

Invariance under CPT has been tested in a variety of experiments
\cite{pdg}.
The tightest bound published to date 
arises from experiments with the neutral kaon system
\cite{expt1},
where the CPT figure of merit 
\beq
r_K \equiv |(m_K - m_{\overline{K}})/m_K|
\label{rK}
\eeq
is known to be smaller than two parts in $10^{18}$.
This remarkable precision is possible because 
neutral-kaon oscillations provide a natural interferometer 
with dimensionless sensitivity 
controlled by the mass difference between
the physical $K_L$ and $K_S$ states:
$|(m_L - m_S)/m_K| \simeq 10^{-14}$.
The quoted precision for $r_K$ 
is thus attained via measurements
with a precision of about one part in $10^4$.

Atomic experiments have also confirmed CPT symmetry.
High-precision comparisons of the anomalous magnetic moments 
of the electron and positron currently provide the
most stringent bounds on CPT violation in lepton systems
\cite{vd}.
Denote the electron and positron $g$ factors by
$g_-$ and $g_+$, respectively. 
Then,
a conventional figure of merit used in these experiments is
\cite{pdg}
\beq
r_g\equiv |(g_- - g_+)/g_{\rm av}|
\quad ,
\label{rg}
\eeq
which is known to be smaller than two parts in $10^{12}$.
The experiments confine isolated single electrons or positrons
in a Penning trap for indefinite periods 
\cite{vd,geo}
and measure their cyclotron and anomaly frequencies
to a precision of better than one part in $10^8$.
These frequencies can be combined to determine $(g-2)$, 
which is of order $10^{-3}$,
and hence to yield the limit on $r_g$.

The figure of merit $r_g$ is poorer than $r_K$ by
about six orders of magnitude,
even though the experimental measurements involved
in the $(g-2)$ experiments are about four orders of
magnitude sharper.
This discrepancy originates in the difference 
between the quantities entering the dimensionless 
figures of merit.
One is a mass (energy) difference
while the other is a coupling difference.
Indeed,
all CPT tests to date have looked for differences 
between particles and antiparticle 
masses, lifetimes, or couplings.
An important limiting factor in comparing 
bounds from various systems
and in establishing new tests
has been the absence of a theoretical framework 
for describing possible CPT violation.

The combination of the theoretical proof of CPT
invariance in conventional field theory and
high-precision tests in experiments 
has triggered investigations of possible CPT violation 
as a candidate signature for
new physics beyond the standard model,
such as string theory
\cite{kp1}.
The current bounds in the kaon system
are close to the scale of suppressed CPT violation
possibly arising in strings
\cite{kp1,kp2},
and new tests in other neutral-meson systems
are feasible with analysis of existing data
or in planned experiments 
\cite{kp2,ckv}.
There are also possible implications for baryogenesis
\cite{bckp}.

Motivated by these ideas,
a theoretical framework for the treatment of 
possible CPT and Lorentz violations at the level of the 
standard SU(3) $\times$ SU(2) $\times$ U(1) model 
has recently been developed
\cite{cptsm}.
Within this framework,
a general CPT- and Lorentz-violating extension 
to the standard model has been presented that appears to
maintain desirable features of the quantum field theory,
including gauge invariance,
naive power-counting renormalizability,
and microscopic causality.
Possible CPT violations are controlled by parameters
with values to be bounded by experiment.

The existence of this model suggests a variety of
experimental approaches to testing CPT
and makes possible a quantitative comparison
of various figures of merit.
In the present work,
we consider a restriction of the model
to quantum electrodynamics to
investigate tests of CPT using the anomalous
magnetic moments of the electron and positron.
In what follows,
we use this model
to show that the conventional figure of merit $r_g$
adopted in $(g-2)$ experiments is a misleading measure
of CPT bounds in lepton systems.
Instead,
an alternative CPT figure of merit is introduced,
and its value within our model is obtained.
A straightforward experimental procedure 
to measure it is proposed,
and an estimate is given of the likely resulting CPT bound.

In the present context,
the dominant CPT-breaking terms from the model act to 
modify the Dirac equation. 
In natural units ($\hbar = c = 1$), the result is 
\beq
\left(i \ga^{\mu} \partial_{\mu} 
- e A_{\mu} \ga^{\mu} - a_{\mu} \ga^{\mu}
- b_{\mu} \ga_5 \ga^{\mu} - m \right) \ps = 0
\quad ,
\label{mdeq}
\eeq
where $\ps$ is the electron-positron field,
$A_\mu$ is the photon field,
$e$ is the electron charge, 
and $m$ is its mass.
The eight quantities $a_\mu$ and $b_\mu$ are 
(small) real constants 
that are invariant under CPT transformations
and act as effective coupling constants.
The standard CPT-transformation properties of $\ps$
can be used to show that the 
terms involving $a_\mu$ and $b_\mu$ break CPT.
These features and Eq.\ \rf{mdeq}
largely suffice to develop the results in the present work.
Various issues concerning other symmetry transformations
(including rotational and boost properties)
and more general extensions of quantum electrodynamics
are treated in Ref.\ \cite{cptsm}
but are not directly relevant here.

In $(g-2)$ experiments,
the leading contributions to the energy spectrum
originate in the particle interaction 
with the constant magnetic field of the Penning trap.
The quadrupole electric field and other fields
produce lesser effects.
Since any possible CPT violation must be small,
it suffices to work within a perturbative framework 
using relativistic quantum mechanics.
The field $\ps$ can thus be regarded as a Dirac wave function
for an electron,
and $A_\mu$ can be treated as a background 
electromagnetic potential.
We denote by $\hat H_0^-$ the conventional Dirac hamiltonian 
operator for an electron in the potential $A_\mu$
for a constant magnetic field,
including an anomaly term.
The exact eigenenergies of $\hat H_0^-$ are the usual Landau levels,
and the eigensolutions can be used as 
the basis for perturbative calculations.
In the presence of the CPT-violating terms
given in Eq.\ \rf{mdeq},
the modified Dirac hamiltonian 
for the electron wave function is 
$\hat H^- = \hat H_0^- + \hat H_{\rm int}^-$,
where
\beq
\hat H_{\rm int}^- = 
a_\mu \ga^0 \ga^\mu - b_\mu \ga_5 \ga^0 \ga^\mu
\quad .
\label{Hint}
\eeq

The wave function for a positron
can be found using charge conjugation.
Typically,
experiments on positrons are performed
in Penning traps with the same magnetic fields as 
used for electron experiments,
with only the electric field changing polarity.
We therefore solve for the positron wave function 
in the same field $A_\mu$ as for the electron.
In the present case,
this implies the usual Dirac hamiltonian $\hat H_0^+$
for a positron is the same as $\hat H_0^-$
except that the coefficient of $A_\mu$ changes sign.
Using the charge conjugation transformation,
the CPT-violating perturbation for the positron
is found to be
\beq
\hat H^+_{\rm int} = - a_\mu \ga^0 \ga^\mu
- b_\mu \ga_5 \ga^0 \ga^\mu
\quad .
\label{Hcint}
\eeq

In investigating CPT-violating effects,
it is unnecessary to include all possible perturbations
that are relevant to $(g-2)$ experiments.
For example,
the effects of the magnetron and axial motions and
the usual higher-order relativistic corrections
are all described within conventional Dirac theory 
and are the same for electrons and positrons.
It therefore suffices to work with the 
electron and positron theories described by $H^\pm_0$.
The point is that all perturbative corrections 
except those involving
$a_\mu$ and $b_\mu$ vanish when the electron and
positron energies are subtracted.
Moreover,
any interactions involving the coupling of $a_\mu$
and $b_\mu$ to other perturbative terms are 
of higher order and therefore can be neglected. 

In what follows,
we denote the relativistic electron and positron Landau-level 
wave functions by $\ps^-_{n,s}$ and $\ps^+_{n,s}$,
respectively.
The corresponding
lowest-order eigenenergies are denoted 
$E^-_{n,s}$ and $E^+_{n,s}$,
where $n=0,1,2,\dots$ labels the level number 
and $s= \pm 1$ labels the spin.
In the electron case
the spin-up and spin-down states form 
two ladders of levels,
for which the spin-down states with given $n=n_0>0$ 
are almost degenerate with the spin-up states with $n=n_0-1$.
The degeneracy is broken due to the anomalous magnetic moment.
A similar situation holds for the positron case,
except that the spin labels are reversed.
The lowest-order cyclotron and anomaly frequencies 
$\om_c^-$ and $\om_a^-$ for the electron
and the corresponding frequencies
$\om_c^+$ and $\om_a^+$ for the positron
can be expressed in terms of the lowest eigenenergies as
\beq
\om_c^\mp = E^\mp_{1,\mp 1} - E^\mp_{0,\mp 1}
\quad , \qquad 
\om_a^\mp = E^\mp_{0,\pm 1} - E^\mp_{1,\mp 1}
\quad .
\label{freq}
\eeq

We orient our coordinate system so that the 
magnetic field $\vec B = B\hat z$ 
lies along the positive $z$ axis,
and we choose the gauge $A^\mu = (0,-yB,0,0)$.
The lowest-order CPT-violating corrections
to the electron energies from $\hat H^-_{\rm int}$ then are 
\beq
\de E^-_{n,\pm 1} = a_0 + a_3 \fr {p_z} {E^-_{n,\pm 1}}
\mp b_3 \left( 1 - \fr { |eB| (2n + 1 \pm 1)}
{E^-_{n,\pm 1} (E^-_{n,\pm 1} + m)} \right)
\mp b_0 \fr {p_z} {E^-_{n,\pm 1}} 
\quad ,
\label{delE}
\eeq
where $p_z\equiv p^3$ is the third component of the momentum.
For the positron,
we find a similar expression but with
the replacements
$a_\mu \to -a_\mu$, 
$E^-_{n,\pm 1} \to E^+_{n,\pm 1}$,
and $\pm 1\to \mp 1$ in the numerator of the third term.

At first sight,
it might appear from these equations
that both $a_\mu$ and $b_\mu$ have physically
observable consequences.
However,
the corrections due to $a_\mu$
correspond to a redefinition 
of the zero of the energy and momentum,
$E \rightarrow E - a^0$ and $\vec p \rightarrow \vec p - \vec a$,
in the dispersion relation for $E^-_{n,s}(\vec p)$.
The corresponding shifts for positrons would have 
opposite signs for $a_\mu$.
Although the electron and positron four-momentum shifts 
are of opposite signs,
they cannot be detected in $(g-2)$ experiments
because the double tower of states in each case 
is shifted so that all level spacings are constant.
The cyclotron and anomaly frequencies remain unchanged
for both cases,
and hence $a_\mu$ has no observable effect
\cite{fn1}.
Without loss of generality,
we can therefore set $a_\mu$ to zero in what follows.

For Penning-trap configurations
typically used in $(g-2)$ experiments,
the axial momentum replaces $p_z$.
Since the energy of the axial motion is 
several orders of magnitude smaller than $E^-_{n,s}$,
the terms in Eq.\ \rf{delE}
involving the product of $b_0$ with $p_z/E^\pm_{n,s}$
can safely be neglected
provided the ratio $b_0/b_3$ is not too large
\cite{fn2}.
For the typical magnetic fields of $B \simeq 5$ T,
$|eB| / m^2 \simeq 10^{-9}$,
so the correction terms involving 
the product of $b_3$ with $|eB|$ can also be ignored.
The dominant CPT-violating contributions 
therefore depend only on $b_3$.
It follows that there are no corrections 
to the cyclotron frequencies,
while the electron and positron anomaly frequencies
shift by $-2 b_3$ and $2 b_3$, 
respectively.
This gives
\beq
\De \om_c \equiv \om_c^- - \om_c^+ = 0 
\quad , \qquad 
\De \om_a \equiv \om_a^- - \om_a^+ = -4 b_3
\quad .
\label{signal}
\eeq
The leading-order signal for CPT breaking 
in Penning-trap $(g-2)$ experiments with fixed magnetic field
is therefore a difference between 
the electron and positron anomaly frequencies.
Note that the signature \rf{signal} for CPT violation
is sensitive only to the spatial components of $\vec b$
in the direction of $\vec B$.
However,
since the relative directions of the two vectors
can be probed experimentally,
for example by changing the orientation of $\vec B$ 
or by performing measurements at different times,
bounds on the different spatial components of $\vec b$
are in principle accessible.

At this point,
we can address the issue of the appropriateness of 
the figure of merit $r_g$ given in Eq.\ \rf{rg} 
as a suitable measure of CPT violation.
Recall that the $g$ factor of an elementary particle
is essentially the strength of the gyromagnetic ratio,
which is the ratio of the magnitudes of the magnetic moment
and the angular momentum.
Conventional quantum electrodynamics for
an electron in a Penning trap predicts 
$(g-2) = 2\om_a/\om_c$,
and CPT invariance predicts $g_- = g_+$.
The latter relation holds to within the 
measurement accuracy of two parts in $10^{12}$.
It therefore appears tempting to use 
the figure of merit $r_g$ of Eq.\ \rf{rg}
as a measure of CPT violation.
However,
within our framework,
CPT is broken without
affecting the electron or positron gyromagnetic ratios.
This means that the theoretical value of $r_g$ is zero 
even though CPT is broken.

One might be tempted to fix this problem by adopting 
as fundamental the conventional experimentally based definition,
$(g_{\rm expt}-2)\equiv 2\om_a/\om_c$,
where $\om_a$ and $\om_c$ are experimental frequencies.
This definition of $g$ would make $r_g$ nonzero
if CPT is violated,
but it would be different from 
the theoretical definition based on the gyromagnetic ratio.
Moreover,
$r_g$ would then depend on the field $B$ 
and might not be well defined.
For example,
our result \rf{signal} means that $r_g$ would become 
$r_g = |\De \om_a / \om^{\rm av}_a| \approx |4b_3/\om^-_a|$,
which diverges in the weak-field limit $B\to 0$.
This provides an explicit counterexample to the thesis
that $r_g$ is a suitable CPT figure of merit.

A more appropriate figure of merit can be introduced  
theoretically in a general context as the ratio of 
a CPT-violating electron-positron energy-level difference 
and the basic energy scale:
\beq
r_e \equiv \left| (\cE^-_{n,s} - \cE^+_{n,-s}) /
\cE^-_{n,s}\right|  
\quad ,
\label{ratio}
\eeq
taken as usual in the weak-field, zero-momentum limit.
Here, 
$\cE^-_{n,s}$ and $\cE^+_{n,s}$ denote energy eigenvalues 
for the full Penning-trap hamiltonians.
Within our particular framework
$\cE^-_{n,s}\to m$ in this limit,
and the difference of energies 
in the numerator becomes half the difference 
between the two measured anomaly frequencies,
$\De \om_a/2 \approx -2 b_3$,
independent of $n$ and $s$.
Thus, in our model the definition \rf{ratio} reduces to  
$r_e = \left| \De\om_a/2m\right| =\left| 2 b_3/m\right|$.
This shows that,
unlike the conventional quantity $r_g$,
the figure of merit $r_e$
is a well defined measure of CPT violation.
Moreover,
since it is a ratio of energies,
it is comparable to the measure $r_K$ 
in Eq.\ \rf{rK} conventionally used
for CPT tests with the neutral-kaon system.

Within the framework of scenarios
involving spontaneous CPT and Lorentz breaking 
from a higher-dimensional fundamental model
such as a string theory
\cite{kp1,kp2,ks},
the natural suppression scale for CPT violation
is the ratio of a light scale $m_l$ to
a large (Planck or compactification) scale $M$.
It is therefore plausible that $r_e \approx m_l/M$.
Some intuition as to the range of possible values for $r_e$
can be found by choosing various values for $m_l$.
If $m_l \approx m$
and taking $M \approx M_{\rm Planck}$,
we find $r_e \simeq 5 \times 10^{-23}$.
If instead $m_l \simeq 250$ GeV,
which is of order of the electroweak scale,
then $r_e \simeq 2 \times 10^{-17}$.

We have seen that 
any existing CPT violation generated by $\vec b$ 
would induce a potentially measurable shift 
between the energy levels of electrons and positrons 
in a Penning trap.
Indeed,
the ratio $r_e$ could be bounded in experiments
using current techniques.
We have investigated several possible experimental procedures 
that could be adopted.
The most effective one would involve taking advantage of the
predicted vanishing of the difference $\De \om_c$
in the electron and positron cyclotron frequencies.
Since $\om_a^\mp$ both depend on the
magnitude of the magnetic field,
it would be important to maintain the calibration of 
$B$ in the measurements of $\De \om_a$.
This could be accomplished by using the equality of the
cyclotron frequencies to verify that the magnetic field
remains the same for both electrons and positrons.
The ratio $r_e$ could then be obtained from
measurements of $\De \om_a$
at equal values of the magnetic field.
These measurements could be repeated using different values
of the magnetic field to verify that $\De \om_a$
is independent of the magnitude $B$ for a fixed orientation
of the field axis.
Since the Penning trap configuration selects the
component of $\vec b$ in the direction of $\vec B$,
an additional check would involve looking for
diurnal variations in the difference $\De \om_a$.

We can estimate the bound on $r_e$ that could be attained.
Suppose the angular anomaly frequencies 
can be measured to an accuracy of approximately 10 Hz. 
This would seem feasible,
for example,
using the line-fitting procedure described in
Ref.\ \cite{vd}.
At the same time,
the equality of the cyclotron frequencies
would have to be maintained
to an accuracy of one part in $10^8$
to account for possible drifts in the magnetic field.
Using Eq.\ \rf{signal}, 
$b_3 = -\De \om_a / 4$.
Assuming no differences in the angular frequency
are observed to this level of precision,
then the bound $|b_3| \lsim 2\times 10^{-15}$ eV can be obtained.
This corresponds to a CPT figure of merit of 
$r_e \lsim 10^{-20}$ in the electron-positron sector.

This estimate suggests a somewhat tighter bound 
for $r_e$ would be attainable 
than that for the corresponding figure of merit $r_K$ 
arising from experiments with the neutral-kaon system.
However, performing the latter tests would continue 
to be essential
because neutral-meson CPT violation is controlled 
by distinct CPT-violating parameters 
appearing in the quark sector.
In any event,
a bound of the estimated magnitude
for $r_e$ in the electron-positron sector 
would be in line with the greater precision 
that is experimentally accessible in a Penning trap using 
measurements of atomic transition frequencies.

\vglue 0.6cm

This work is supported in part by the U.S.\ D.O.E.\
under grant number DE-FG02-91ER40661 and by the N.S.F.\ 
under grant number PHY-9503756.
One of us (R.B.) thanks Indiana University for hospitality 
and Colby College for a Science Division grant.

\vglue 0.6cm


\begin{thebibliography}{xx}

\bibitem{pct}
See, for example,
R.F. Streater and A.S. Wightman,
{\it PCT, Spin, Statistics, and All That}
(Benjamin Cummings, London, 1964).

\bibitem{pdg}
See, for example,
R.M. Barnett {\it et al.},
Review of Particle Properties,
Phys. Rev. D {\bf 54} (1996) 1.

\bibitem{expt1}
L.K. Gibbons et al., Phys. Rev. D {\bf 55} (1997) 6625;
B. Schwingenheuer et al., Phys. Rev. Lett. {\bf 74} (1995) 4376;
R. Carosi et al., Phys. Lett. B {\bf 237} (1990) 303.

\bibitem{vd}
R.S. Van Dyck, Jr., P.B. Schwinberg, and H.G. Dehmelt, 
Phys. Rev. Lett. {\bf 59} (1987) 26;
Phys. Rev. D {\bf 34} (1986) 722.

\bibitem{geo}
L.S. Brown and G. Gabrielse,
Rev. Mod. Phys. {\bf 58} (1986) 233.

\bibitem{kp1}
V.A. Kosteleck\'y and R. Potting,
Nucl. Phys. B {\bf 359} (1991) 545;
Phys. Lett. B {\bf 381} (1996) 389.

\bibitem{kp2}
V.A. Kosteleck\'y and R. Potting,
Phys. Rev. D {\bf 51} (1995) 3923.
See also V.A. Kosteleck\'y and R. Potting,
in D.B. Cline, ed.,
{\it Gamma Ray--Neutrino Cosmology and Planck Scale Physics} \rm
(World Scientific, Singapore, 1993)
(hep-th/9211116).

\bibitem{ckv}
D. Colladay and V. A. Kosteleck\'y,
Phys. Lett. B {\bf 344} (1995) 259;
Phys. Rev. D {\bf 52} (1995) 6224;
V.A. Kosteleck\'y and R. Van Kooten,
Phys. Rev. D {\bf 54} (1996) 5585.

\bibitem{bckp}
O. Bertolami et al.,
Phys. Lett. B {\bf 395}, 178 (1997).

\bibitem{cptsm}
D. Colladay and V.A. Kosteleck\'y,
Phys. Rev. D {\bf 55} (1997) 6760.

\bibitem{fn1}
Note that the coefficients $a_\mu$ \it cannot \rm
be removed by a gauge transformation 
on the electron and photon fields
because the full lagrangian,
including the term involving $a_\mu$,
is gauge invariant.
Instead,
in any theory with only one fermion field,
$a_\mu$ can be eliminated  
by a field redefinition coupled with a background gauge change.
This is equivalent to the redefinition of the energy and momentum
as described in the text.
These results have been proved in a general context in
\cite{cptsm},
where it was shown that interactions between two fermion fields 
can produce effects depending on the difference between 
couplings in two distinct terms of this type.
This is irrelevant for
separately trapped electrons or positrons.
Incidentally,
the effects of the $a_\mu$ term 
for the electron-positron field are unobservable 
even in Penning-trap experiments 
determining the electron-positron mass ratio 
\cite{schwin81},
since the separate cyclotron-frequency
measurements for the electron and positron
remain unaffected by the $a_\mu$ coupling.

\bibitem{schwin81}
P.B. Schwinberg, R.S. Van Dyck, Jr., and H.G. Dehmelt,
Phys. Lett. A {\bf 81} (1981) 119.

\bibitem{fn2}
This always holds if $b_\mu$ is spacelike or lightlike,
and it also holds for a large range of timelike $b_\mu$.
In fact, 
a stronger result can be obtained.
The axial motion can be incorporated into the analysis
via a Foldy-Wouthuysen diagonalization
of the full relativistic hamiltonian 
allowing for both magnetic and electric fields.
The resulting hamiltonian appears more complicated 
than the expressions in the text;
for example,
involving an operator momentum $\vec p$
rather than the constant linear momentum $p_z$ of Eq.\ \rf{delE}.
It can be shown by explicit calculation or by
using P symmetry that terms 
linear in $b_0$ produce no contributions
to the energy levels $E^\pm_{n,s}$.
This means experiments are sensitive only to $b_0^2$ at best,
and the corresponding limits on $b_0$ would  
therefore be of more limited interest.

\bibitem{ks}
V.A. Kosteleck\'y and S. Samuel,
Phys. Rev. Lett. {\bf 63} (1989) 224;
ibid. {\bf 66} (1991) 1811;
Phys. Rev. D {\bf 39} (1989) 683;
ibid. {\bf 40} (1989) 1886.

\end{thebibliography}
\end{document}